\documentclass[a4paper]{article}
\RequirePackage[english]{babel}
\RequirePackage[latin1]{inputenc}
\RequirePackage[T1]{fontenc}
\RequirePackage{mathrsfs}
\RequirePackage{amsmath}
\RequirePackage{amssymb}
\RequirePackage{amsbsy}
\RequirePackage{bm}
\begin{document}
\title{\bf{A Torsional Model of Leptons}}
\author{Luca Fabbri\\ 
\footnotesize DIME Sez. Metodi e Modelli Matematici, Universit\`{a} di Genova\\
\footnotesize INFN \& Dipartimento di Fisica, Universit\`{a} di Bologna}
\date{}
\maketitle
\begin{abstract}
Quite recently it was shown that torsion induces interactions among leptons that are identical to the weak interactions of leptons in the Weinberg standard model, if it is in terms of leptonic bound states that the bosonic sector is built; here we obtain the partially conserved axial currents showing that they are the same of the standard model, if the composite mediators have specific mass relationships: we show that their masses are indeed the measured ones, if reasonable approximations are taken.
\end{abstract}
\section*{Introduction}
Very recently, the LHC experiments ATLAS and CMS have announced the discovery of a new particle at a mass that is compatible with that of the expected Higgs field; although this new boson has not been identified with the Higgs boson yet, high are the chances that this will happen: if this will indeed be the case, the last empty slot of the Weinberg leptonic standard model will be filled, and unless indications for new physics will emerge, the standard model might be thought as complete. In talking about completeness of the standard model of leptons, we mean that all predictions of the standard model have been confirmed and vice versa; what is instead not meant is the theoretical structure of the standard model itself. For instance, although the particle content of the standard model may be matched by the one observed in accelerators, it may well be that at the present energies we may recognize as fundamental fields that at higher energies might be composite. Despite this momentous experimental improvement, the present situation does not radically change what has ever been thought about the Higgs boson: that is, although people always had a strong feeling that it would have been found, nevertheless there has also been a generalized tendency to replace the fundamental Higgs boson with composite states of other more fundamental fields. One of the first attempts in this direction was due to Weinberg himself \cite{w}, modelling leptons in terms of the Nambu-Jona--Lasinio theory, the replacement of the fundamental with a composite Higgs scalar implying the replacement of the spontaneous with a dynamic symmetry breaking mechanism. Therefore, from the Weinberg standard model of leptons and gauge fields with spontaneous symmetry breaking \cite{Fabbri:2009ta}, a model of leptons with gauge fields in dynamical symmetry breaking can be obtained.

From this point on, a further degree of simplicity can be obtained if beside the scalar field also the mediator fields are composite in terms of something that is more fundamental: the motivation that pushes us to pursue this goal is that when modelling leptons, for which the balance of right-handed and left-handed components is not even, in terms of Nambu-Jona--Lasinio potentials, which can be rearranged as either scalar or vector fields, the lack of balance between the two chiral projections favours the emergence beside the usual scalar fields of the vector fields. This would provide the mould to form beside a composite Higgs also the composite weak mediators, so that the whole bosonic sector will emerge as a composite state of something more fundamental, which can only be the leptonic fields, held together by Nambu-Jona--Lasinio interactions, whose lack of chirality would imply that there will be no need to break an initial symmetry because there will be no symmetry that would ever be present \cite{Fabbri:2010ux}.

As a final remark we notice that Nambu-Jona--Lasinio equations have a natural geometrical interpretation: in considering the most general geometric background in which Cartan torsion tensor is present in the most general covariant derivative, the Dirac field equation with torsion can be worked out to be equivalent to the Dirac equation without torsion but with self-interacting quartic potentials having the form of Nambu-Jona--Lasinio potentials \cite{Fabbri:2011kq}.
\section{Torsionally-Interacting Lepton Fields}
As we have already discussed in the introduction, it is easy to summarize the layout of the present construction: the most general geometric background will be considered in which torsion accounts for self-interactions in the Dirac field equation, making it the Nambu-Jona--Lasinio field equation \cite{Fabbri:2011kq}; when this field equation is applied to the case of leptons, the unbalanced left-handed components will give rise beside the usual scalar fields also to vector fields, which can be used for modelling beside the Higgs also the mediators in a construction in which the whole bosonic sector would be composite \cite{Fabbri:2010ux}. In doing so, we are going to obtain a system of torsionally-interacting leptons that has to be compared to that of the standard model of leptons \cite{Fabbri:2009ta}: we will see that in the two cases, the field equations for leptons are essentially identical if the mediators are bound-states of leptons, and we will employ the partially conserved axial currents to evaluate the masses of the composite mediators seeing that they are equal to the measured ones if some reasonable approximations are taken.

We begin by considering the Clifford gamma matrices $\gamma_{\mu}$ and the parity-odd gamma matrix $\gamma=i\gamma^{0}\gamma^{1}\gamma^{2}\gamma^{3}$ in terms of which a pair of leptons $e$ with charge $q$ and mass $m$ and $\nu$ neutral and massless can be decomposed into their chiral projections $2e_{L}\!\!=\!(\mathbb{I}-\gamma)e$ and $2e_{R}\!\!=\!(\mathbb{I}+\gamma)e$ with $\nu_{L}\!=\!\nu$ and $\nu_{R}\!=\!0$ and this will constitute our matter content, with covariant derivative $\nabla_{\mu}$ and torsional contributions with coupling constant tuned to the Fermi constant $G_{F}$ so that
\begin{eqnarray}
&i\gamma^{\mu}\nabla_{\mu}e
-\frac{G_{F}}{\sqrt{2}}\overline{e}\gamma_{\mu}e\gamma^{\mu}e
-\frac{G_{F}}{\sqrt{2}}\overline{\nu}\gamma_{\mu}\nu\gamma^{\mu}\gamma e-me=0
\label{1}\\
&i\gamma^{\mu}\nabla_{\mu}\nu
-\frac{G_{F}}{\sqrt{2}}\overline{e}\gamma_{\mu}\gamma e\gamma^{\mu}\nu=0
\label{2}
\end{eqnarray}
is the general system of field equations \cite{Fabbri:2011kq}; by employing Fierz identities and by introducing for generality the parameter $g$ such that the electronic charge can be written as $q=g\sin{\theta}$ in terms of the Weinberg angle and the Yukawa constant $Y$ such that the electronic mass can be written as $m=Yv$ in terms of the constant $v=\frac{1}{\sqrt{\sqrt{8}G_{F}}}$ then the field equations (\ref{1}-\ref{2}) can be rewritten as
\begin{eqnarray}
\nonumber
&i\gamma^{\mu}\nabla_{\mu}e+G_{F}q\tan{\theta}\frac{\sqrt{2}\cos{\theta}}{g\left[2\sin{\theta}\right)^{2}}
\left[2(\overline{e}_{L}\gamma_{\mu}e_{L}-\overline{\nu}\gamma_{\mu}\nu)
-(2\sin{\theta})^{2}\overline{e}\gamma_{\mu}e\right]\gamma^{\mu}e-\\
\nonumber
&-\frac{gG_{F}}{2\cos{\theta}}\frac{\sqrt{2}\cos{\theta}}{g\left(2\sin{\theta}\right)^{2}}
\left[2(\overline{e}_{L}\gamma_{\mu}e_{L}-\overline{\nu}\gamma_{\mu}\nu)
-(2\sin{\theta})^{2}\overline{e}\gamma_{\mu}e\right]\gamma^{\mu}e_{L}+\\
\nonumber
&+\frac{gG_{F}}{\sqrt{2}}\frac{\left[1-(2\sin{\theta})^{2}\right]}{g\left(2\sin{\theta}\right)^{2}}\left(2\overline{\nu}\gamma_{\mu}e_{L}\right)\gamma^{\mu}\nu-\\
&-Y\frac{G_{F}}{Y}\sqrt{2}(\cos{\theta})^{2}\overline{e}ee
+G_{F}\sqrt{2}(\cos{\theta})^{2}\overline{e}\gamma e\gamma e-me=0
\label{electronic}\\
\nonumber
&i\gamma^{\mu}\nabla_{\mu}\nu+\frac{gG_{F}}{2\cos{\theta}}\frac{\sqrt{2}\cos{\theta}}{g\left(2\sin{\theta}\right)^{2}}
\left[2(\overline{e}_{L}\gamma_{\mu}e_{L}-\overline{\nu}\gamma_{\mu}\nu)
-(2\sin{\theta})^{2}\overline{e}\gamma_{\mu}e\right]\gamma^{\mu}\nu+\\
&+\frac{gG_{F}}{\sqrt{2}}\frac{\left[1-(2\sin{\theta})^{2}\right]}{g\left(2\sin{\theta}\right)^{2}}\left(2\overline{e}_{L}\gamma_{\mu}\nu\right)\gamma^{\mu}e_{L}=0
\label{neutrinic}
\end{eqnarray}
where the torsionally-induced chiral interactions between left-handed and right-handed projections of all fields involved have been cast into scalar electronic self-interactions plus vector currents between electron and neutrino as mutual interactions among leptons \cite{Fabbri:2010ux}: we may rewrite field equations (\ref{electronic}-\ref{neutrinic}) as
\begin{eqnarray}
\nonumber
&i\gamma^{\mu}\nabla_{\mu}e+G_{F}\sqrt{2}(\cos{\theta})^{2}\overline{e}\gamma e\gamma e
+q\tan{\theta}Z_{\mu}\gamma^{\mu}e-\\
&-\frac{g}{2\cos{\theta}}Z_{\mu}\gamma^{\mu}e_{L}
+\frac{g}{\sqrt{2}}W^{*}_{\mu}\gamma^{\mu}\nu-YHe-me=0
\label{electron}\\
&i\gamma^{\mu}\nabla_{\mu}\nu+\frac{g}{2\cos{\theta}}Z_{\mu}\gamma^{\mu}\nu
+\frac{g}{\sqrt{2}}W_{\mu}\gamma^{\mu}e_{L}=0
\label{neutrino}
\end{eqnarray}
upon definition of the following bosons, being them scalars
\begin{eqnarray}
&H=\frac{G_{F}}{Y}\sqrt{2}(\cos{\theta})^{2}\overline{e}e
\label{Higgs}
\end{eqnarray}
or vectors
\begin{eqnarray}
&Z_{\mu}=G_{F}\frac{\sqrt{2}\cos{\theta}}{g\left(2\sin{\theta}\right)^{2}}
\left[2\left(\overline{e}_{L}\gamma_{\mu}e_{L}-\overline{\nu}\gamma_{\mu}\nu\right)
-(2\sin{\theta})^{2}\overline{e}\gamma_{\mu}e\right]
\label{neutral}\\
&W_{\mu}=G_{F}\frac{\left[1-(2\sin{\theta})^{2}\right]}{g\left(2\sin{\theta}\right)^{2}}
\left(2\overline{e}_{L}\gamma_{\mu}\nu\right)
\label{charged}
\end{eqnarray}
which can be compared to the system of field equations of the Weinberg standard model for the leptonic fields \cite{Fabbri:2009ta} given by the following expressions
\begin{eqnarray}
&i\gamma^{\mu}\nabla_{\mu}e+q\tan{\theta}Z_{\mu}\gamma^{\mu}e
-\frac{g}{2\cos{\theta}}Z_{\mu}\gamma^{\mu}e_{L}
+\frac{g}{\sqrt{2}}W_{\mu}^{\ast}\gamma^{\mu}\nu-YHe-me=0
\label{electronSM}\\
&i\gamma^{\mu}\nabla_{\mu}\nu
+\frac{g}{2\cos{\theta}}Z_{\mu}\gamma^{\mu}\nu
+\frac{g}{\sqrt{2}}W_{\mu}\gamma^{\mu}e_{L}=0
\label{neutrinoSM}
\end{eqnarray}
in terms of the fields $H$ and also $Z_{\mu}$ and $W_{\mu}$ which are now assumed to be intrinsically structureless: after their comparison we see that the respective field equations for leptons (\ref{electron}-\ref{neutrino}) and (\ref{electronSM}-\ref{neutrinoSM}) are essentially identical, in which the scalar field and the leptonic currents (\ref{Higgs}) and (\ref{neutral}-\ref{charged}) play the role of the composite Higgs boson and weak mediators that are fundamental in the standard model.

\subsection{Partially-Conserved Axial Currents}
So far we have obtained a model in which torsion binds leptons together to form scalars and currents and where there is no need to brake symmetries because the configuration is never symmetric; on the other hand, a torsionless model of leptons is built in terms of gauge symmetries eventually broken by the Higgs vacuum expectation: as it stands, it is quite astonishing that these two different models are in fact capable of reproducing field equations for leptons that are indeed identical if the bosonic sector is given in terms of composite leptons.

Now because the bosons are bound-states of leptons, of which we know the dynamics, it is possible to obtain the dynamical properties of the bosonic sector as well, so that by naming $G_{F}\overline{e}e=2\sqrt{2}m\varepsilon$ one can compute the divergences of the leptonic currents (\ref{neutral}-\ref{charged}) obtaining the partially conserved axial currents
\begin{eqnarray}
&\frac{gv(2\sin{\theta})^{2}\left[\left(1+\frac{H}{v}\right)\nabla_{\mu}Z^{\mu}
+2Z^{\mu}\nabla_{\mu}\frac{H}{v}\right]}
{\sqrt{2}\cos{\theta}+\frac{\varepsilon}{\sqrt{2}}(2\cos{\theta})^{3}
-\frac{g\left(2\sin{2\theta}\right)^{2}}{G_{F}}
\frac{Z^{\mu}}{m i\overline{e}\gamma e}\partial_{\mu}\varepsilon}
\!=\!-\sqrt{\!\sqrt{2}G_{F}}m\left(i\overline{e}\gamma e\right)
\label{conservedneutral}\\
&\frac{gv\left[\left(1+\frac{H}{v}\right) \left(\nabla_{\mu}W^{\mu}+iq\tan{\theta}Z^{\mu}W_{\mu}\right)
+2W^{\mu}\nabla_{\mu}\frac{H}{v}\right]}
{\sqrt{2}\left[1-(2\sin{\theta})^{2}\right]
\left[1+\varepsilon(2\cos{\theta})^{2}\right]\left[\varepsilon
-\frac{1}{2(2\sin{\theta})^{2}}\right]+\frac{g(2\cos{\theta})^{2}}{\sqrt{2}G_{F}}
\frac{W^{\mu}}{m i\overline{e}\gamma\nu}\partial_{\mu}\varepsilon}
\!\!=\!\!\sqrt{\!\!\sqrt{8}G_{F}}m\!\left(i\overline{e}\gamma\nu\right)
\label{conservedcharged}
\end{eqnarray}
as it can be obtained through a direct calculation \cite{Fabbri:2010ux} whereas the structureless boson given by $Z_{\mu}$ and $W_{\mu}$ have partially conserved axial currents
\begin{eqnarray}
&m_{Z}\left[\left(1+\frac{H}{v}\right)\!\nabla_{\mu}Z^{\mu}\!
+\!2Z^{\mu}\nabla_{\mu}\frac{H}{v}\right]
\!=\!-\sqrt{\sqrt{2}G_{F}}m\left(i\overline{e}\gamma e\right)
\label{constraintneutral}\\
&m_{W}\left[\left(1+\frac{H}{v}\right)\!\left(\nabla_{\mu}W^{\mu}\!+\!iq\tan{\theta}Z^{\mu}W_{\mu}\right)\!+\!2W^{\mu}\nabla_{\mu}\frac{H}{v}\right]\!
=\!\sqrt{\sqrt{8}G_{F}}m\left(i\overline{e}\gamma\nu\right)
\label{constraintcharged}
\end{eqnarray}
as it can be checked in \cite{Fabbri:2009ta}, and once again they can compared: after the comparison has been performed, (\ref{conservedneutral}-\ref{conservedcharged}) and (\ref{constraintneutral}-\ref{constraintcharged}) are identical if the relationships
\begin{eqnarray}
&m_{Z}\!=\!\frac{gv(2\sin{\theta})^{2}}
{\sqrt{2}\cos{\theta}+\frac{\varepsilon}{\sqrt{2}}(2\cos{\theta})^{3}
-\frac{g\left(2\sin{2\theta}\right)^{2}}{G_{F}}
\frac{Z^{\mu}}{m i\overline{e}\gamma e}\partial_{\mu}\varepsilon}
\label{massZ}\\
&m_{W}\!=\!\frac{gv}{\sqrt{2}\left[1-(2\sin{\theta})^{2}\right]
\left[1+\varepsilon(2\cos{\theta})^{2}\right]\left[\varepsilon
-\frac{1}{2(2\sin{\theta})^{2}}\right]+\frac{g(2\cos{\theta})^{2}}{\sqrt{2}G_{F}}
\frac{W^{\mu}}{m i\overline{e}\gamma\nu}\partial_{\mu}\varepsilon}
\label{massW}
\end{eqnarray}
giving the composite mediators mass relationships are taken to be valid.

\subsubsection{Composite-Mediators Mass Relationships}
As it is clear from the previous analysis, the model in which torsion binds leptons together to form leptonic currents and the one in which a torsionless model of leptons is built with gauge fields give the same partially conserved axial currents if the composite mediators mass relationships have a specific form, which in fact happens to be valid if some reasonable approximations are considered.

Considering the mass relationships (\ref{massZ}-\ref{massW}) we may proceed now to their evaluation, for which the first point we have to notice is that, quite in general, the leptonic currents are proportional to the momentum they transfer through the leptonic propagation given by $m\overline{\psi}\gamma_{\mu}\psi\approx P_{\mu}\overline{\psi}\psi$ while in momentum space the derivatives of a field are proportional to the momentum carried along the propagation of the field itself given by $\partial_{\mu}(\overline{\psi}\psi)\approx P_{\mu}\overline{\psi}\psi$ as it is known, so that by setting the difference $\frac{P^{2}}{m^{2}}-1=\delta$ and parametrizing the measured value of the Weinberg angle $(2\sin{\theta})^{2}\!\approx\!0.93$ as $(2\sin{\theta})^{2}\!\equiv\!1\!-\!\alpha$ with $\alpha\approx7\cdot10^{-2}$ we get
\begin{eqnarray}
&m_{Z}\approx\frac{gv(1-\alpha)}{\sqrt{2}\cos{\theta}\left(1-3\varepsilon\delta\right)}
\label{massZapprox}\\
&m_{W}\approx\frac{gv}{\sqrt{2}\alpha\left(3\varepsilon^{2}+\varepsilon-\frac{1}{2}
+\frac{3}{2}\varepsilon\delta\right)}
\label{massWapprox}
\end{eqnarray}
which coincide to the known values
\begin{eqnarray}
&m_{Z}\approx\frac{gv}{\sqrt{2}\cos{\theta}}\\
&m_{W}\approx\frac{gv}{\sqrt{2}}
\end{eqnarray}
whenever the relationship $\varepsilon^{-6}\approx\delta^{2}\approx\alpha^{3}\approx10^{-6}$ among the order of magnitude of the different parameters happens to be true: notice that, at the weak interaction scales, we have $\overline{e}e\approx10^{-54}L^{-3}_{\mathrm{Planck}}$ yielding $\varepsilon\approx10$ while, although for free particles delta vanishes, for interacting particles it is not necessarily equal to zero, so that it is reasonable to assume it to be small, showing that the approximation we assumed, despite it was rough, was rather reasonable.
\section*{Conclusion}
In the present paper, we have taken into account the general theory formerly provided in \cite{Fabbri:2011kq} in order to study the torsionally-interacting system of two leptonic fields applying the method outlined in \cite{Fabbri:2010ux} and thus obtaining results that had to eventually be compared to the results reviewed in \cite{Fabbri:2009ta}: we have first of all considered that the most general system of field equations has potentials of self-interactions with an unknown coupling constant that can be chosen to be the Fermi constant while the self-interactions among leptons can be rearranged in the form of the weak interactions for those leptons, if the bosonic sector is given in terms of composite leptonic fields; in the present model the bosons, whether they are scalars or currents, must display internal structure at scales in which it is possible to probe their compositeness. At the scales of weak forces, the composite mediators behave as if they were fundamental, and therefore their properties look like those of the weak bosons: because the partially conserved axial currents are what makes it possible to evaluate the mediators masses, we have calculated the partially conserved axial currents through which we have evaluated the composite mediators mass relationships. Eventually, we have seen that their values are in fact the measured ones, if some reasonable approximation is taken into account. To push further the precision numerical simulations are clearly needed, although we will postpone this task to a following paper.

Now it would of course be na\"{\i}f to think that the present approach could fit data successfully as the standard model does, but if that were to be the case then the present approach should be considered a better explanation of nature, because it is more economic than the standard model: in fact we have that in the standard model, or some of its simplest extensions \cite{w}, the phenomenology is obtained after the spontaneous, or in some cases dynamical, breaking of an initial symmetry, where the breaking is obtained by assuming the existence of the Higgs boson with specific interactions, and the initial symmetry is due to the masslessness of the leptons; in the present model instead there is no symmetry from the beginning because of the massiveness of leptons, and there is the torsion tensor. In the standard model, some may argue that the Higgs sector has to be added in order to get the most general action possible, but in the present model, the very same argument may be used to justify torsion since it is present in the most general geometric background; in the present model, some may argue that the particle masses must be assigned by hand and that is inelegant, but nevertheless in the standard model, the very same argument may be used to acknowledge that the Yukawa couplings are inelegant in exactly the same way. Apart from the critics of elegance for the approaches of these two models, there are less terms to be added in the action and fewer parameters to be adjusted in the resulting potentials here than in the standard model, and this undoubtedly makes the present model better than the standard model itself in terms of economy of assumptions. And of course, this fact is even more dramatic for any of the non-torsional common extensions of the standard model.

Conversely, approaches in which extended gravities are used to study the standard model have been developed, such as \cite{Capozziello:2011au,Capozziello:2011et} or alternative approaches based on torsion as a massive field in \cite{Belyaev:2007fn}, where it is also discussed the possibility to detect torsion at the LHC if some reasonable additional conditions on the extended dynamics are met; an interesting alternative on the role of extended gravity is the one given in reference \cite{z} where torsion at high-energies is taken into account in order to replace technicolor. More alternative models are discussed in some of the references of the aforementioned papers.

With the present paper, our aim was to present one of the possible alternatives of the standard model that was not based on generalizing the particle sector but the underlying geometry, showing that in fact interesting results can actually be obtained in this case too, although of course much more should be done to make this model more reliable.


\begin{thebibliography}{12}
\bibitem{w}
S.~Weinberg,
\textit{Phys. Rev.} \textbf{D13}, 974 (1976).
\bibitem{Fabbri:2009ta} 
L.~Fabbri,
\textit{Mod. Phys. Lett. A} \textbf{26}, 2091 (2011).
\bibitem{Fabbri:2010ux} 
L.~Fabbri,
\textit{Int. J. Theor. Phys.} \textbf{50}, 3616 (2011).
\bibitem{Fabbri:2011kq}
L.~Fabbri,
arXiv:1108.3046 [gr-qc].
\bibitem{Capozziello:2011au} 
S.~Capozziello, G.~Basini, M.~De Laurentis,\\
\textit{Eur. Phys. J. C} \textbf{71}, 1679 (2011).
\bibitem{Capozziello:2011et} 
S.~Capozziello, M.~De Laurentis,
\textit{Phys. Rept.} \textbf{509}, 167 (2011).
\bibitem{Belyaev:2007fn}
A.~S.~Belyaev, I.~L.~Shapiro, M.~A.~B.~do Vale,\\
\textit{Phys. Rev. D} \textbf{75}, 034014 (2007).
\bibitem{z}
M.~A.~Zubkov,
\textit{Mod. Phys. Lett. A} \textbf{25}, 2885 (2010).
\end{thebibliography}
\end{document}